\documentclass[fleqn,usenatbib]{mnras}

\usepackage[T1]{fontenc}

\DeclareRobustCommand{\VAN}[3]{#2}
\let\VANthebibliography\thebibliography
\def\thebibliography{\DeclareRobustCommand{\VAN}[3]{##3}\VANthebibliography}

\usepackage{physics} 
\usepackage{commath} 

\usepackage{graphicx}	
\usepackage{amsmath}	
\usepackage{amssymb}	

\title[Galaxies as clocks and the universal expansion]{On the use of galaxies as clocks and the universal expansion}

\author[A. Ahlström Kjerrgren and E. Mörtsell]{
Anders Ahlström Kjerrgren$^{1}$\thanks{E-mail: kjerr@kth.se}
and Edvard Mörtsell$^{2}$\thanks{E-mail: edvard@fysik.su.se}
\\
$^{1}$KTH Royal Institute of Technology, SE 106 91 Stockholm, Sweden\\
$^{2}$Oskar Klein Centre, Department of Physics, Stockholm University, SE 106 91 Stockholm, Sweden
}

\date{Accepted XXX. Received YYY; in original form ZZZ}

\pubyear{2022}

\begin{document}
\label{firstpage}
\pagerange{\pageref{firstpage}--\pageref{lastpage}}
\maketitle

\begin{abstract}
We set out to rederive the 8 Hubble parameter values obtained from estimated relative galaxy ages by Simon et al.~[Physical Review D, 71, 123001 (2005)]. We find that to obtain the level of precision claimed in $H(z)$, unrealistically small galaxy age uncertainties have to be assumed. Also, some parameter values will be correlated. In our analysis we find that the uncertainties in the Hubble parameter values are significantly larger when 8 independent $H(z)$ are obtained using Monte Carlo sampling. Smaller uncertainties can be obtained using Gaussian processes, but at the cost of strongly correlated results. We do not obtain any useful constraints on the Hubble parameter from the galaxy data employed.
\end{abstract}




\section{Introduction}\label{sec:intro}
In 2005 the paper titled ``Constraints on the redshift dependence of the dark energy potential'' was published in Physical Review D \citep{simon2005constraints}. In section IV.A of this paper 8 independent values of the Hubble parameter are obtained at a large range of redshifts to a relatively high degree of precision. These values are being used to test other aspects of cosmology, see for instance \cite{gomez2020update, odintsov2020testing, chen2017determining, melia2018model, gomez2018h0}. Some papers use the results, but cite compilations which include the results from \cite{simon2005constraints} instead; examples include \cite{renzi2020look} who cited \cite{stern2010cosmic}, and \cite{bonilla2021measurements} who cited \cite{moresco20166}. 

The purpose of this work is to investigate the reliability of the resulting values of the Hubble parameter. The practical implementation of the method used in \cite{simon2005constraints} is not trivial. First, it requires determining the ages of galaxies accurately and precisely to obtain the dependence on the redshift, $t(z)$. The second step is to infer the derivative $\dif t/\dif z$ of this relation, a task more difficult than inferring the function values in the sense that it, in general, leads to larger uncertainties. Assuming that the first step is done flawlessly, we show that the results in \cite{simon2005constraints} assumes unrealistically small galaxy age uncertainties due to the difficulty in making precise inferences of derivatives. 8 values of the Hubble parameter are obtained, and therefore 8 derivatives are computed, with uncertainties between 10 and 20$\%$ using 32 galaxies, indicating that age uncertainties much smaller than the reported 12\% have been assumed.\footnote{This work is based on the master's thesis \cite{kjerrgren2021galaxies}, written by Anders Ahlström Kjerrgren and supervised by Edvard Mörtsell.}

\section{The Hubble parameter from galaxy data}
It was proposed by \cite{jimenez2002constraining} that if passively evolving galaxies at different redshifts are considered the Hubble parameter can be obtained by computing the age difference of the galaxies. The equation
\begin{equation}\label{eq:HubbleparaFromDerivative}
    H(z) = - \frac{1}{1+z} \dv{z}{t}
\end{equation}
relates the relevant quantities. The derivative $\dif z / \dif t$ is taken with respect to the cosmic time, however if the galaxies considered are born at the same cosmic time the derivative eliminates the constant offset and galaxy ages can thus be used as a proxy for the cosmic time. The derivative is then further approximated as the ratio $\Delta z/\Delta t$.

\begin{figure*}
    \centering
    \includegraphics[width=0.95\textwidth]{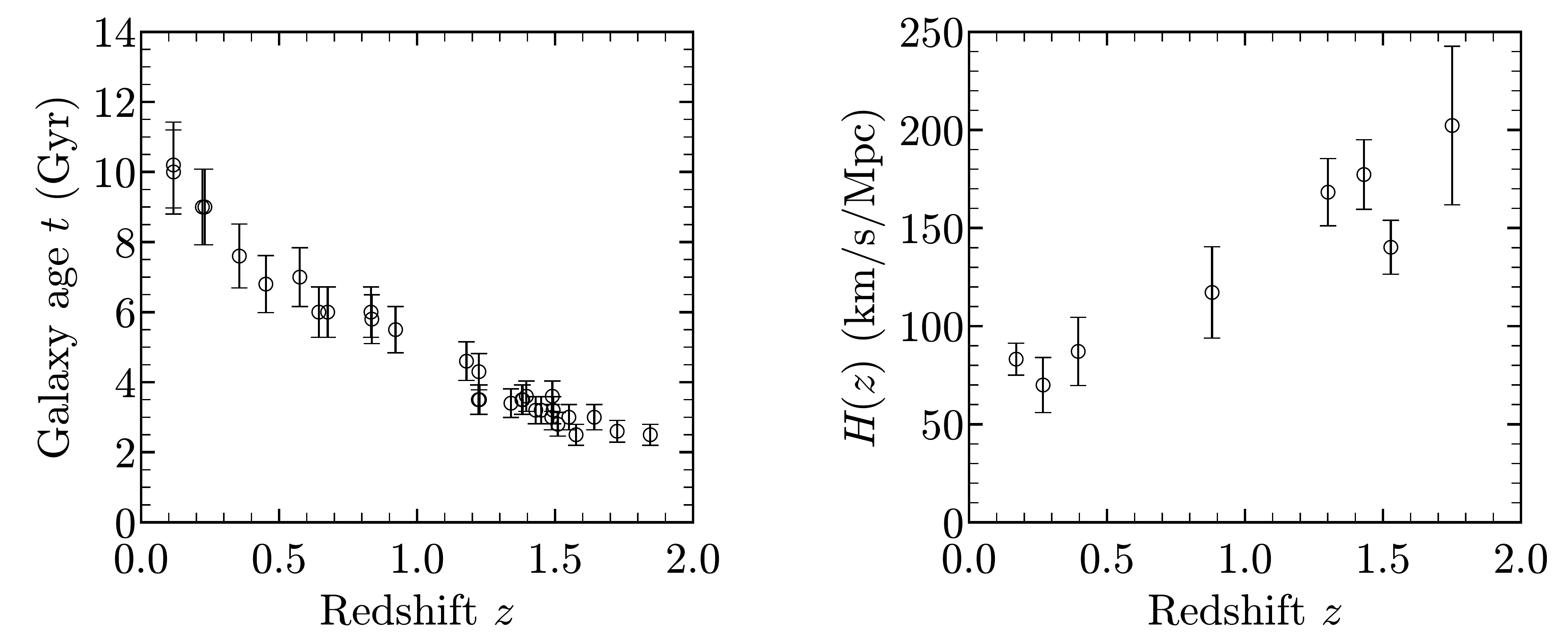}
    \caption{Data and results in \protect\cite{simon2005constraints}. Error bars correspond to 1-$\sigma$ uncertainties. \textbf{Left}: The galaxy age and redshift data. All age uncertainties are 12\%, as given to \protect\cite{samushia2010constraints} by the authors in a private communication. \textbf{Right}: The Hubble parameter values.}
    \label{fig:simonEtAlDataAndResults}
\end{figure*}

The method requires the identification of the same type of galaxies which at lower redshifts are older versions of those at higher redshifts \citep{weinberg2013observational}. Identifying such galaxies yields a relation between galaxy age and redshift. This relation is visible in the data used by \cite{simon2005constraints}, shown in the left panel of Fig.~\ref{fig:simonEtAlDataAndResults}.

By binning the data and using equation \eqref{eq:HubbleparaFromDerivative} 8 determinations of the Hubble parameter are obtained by \cite{simon2005constraints}, shown in the right panel of Fig.~\ref{fig:simonEtAlDataAndResults}, with 10 or 20\% 1-$\sigma$ uncertainties. 

\section{Attempt at reproducing the results}
We first attempt to reproduce the results as close as possible. For clarity we focus on a few select $H(z)$: the ones at the lowest, second lowest, and highest redshift. See appendix \ref{appendix:Reproducing} for a full attempt at reproducing the results. 

With the four galaxies at the lowest redshifts one obtains the effective redshift ${z\simeq0.17}$ corresponding to the Hubble parameter value at the lowest redshift. Determining $H(z)$ requires the determination of the derivative $\dif z / \dif t$. By fitting a straight line to the data using $\chi^2$ minimisation we identify the slope as the reciprocal derivative $\dif t / \dif z$. Combining this with equation \eqref{eq:HubbleparaFromDerivative} yields the same central value as \cite{simon2005constraints}, namely $H(z\simeq 0.17) \simeq 83$ km/s/Mpc. However, using 12\% uncertainties in the galaxy ages results in $> 100\%$ uncertainty in the Hubble parameter. It is also obtained when computing a weighted average of the first and second galaxy, and the third and fourth galaxy, and then calculating $H(z)$ with $\Delta z/\Delta t$ using these two weighted averages and regular uncertainty propagation. To obtain 10\% uncertainty in $H(z)$, as \cite{simon2005constraints}, the age uncertainties must be lowered to $\simeq 1.2\%$, i.e.~a factor of ten.

The Hubble parameter at the second lowest redshift must be computed using the third, fourth, and fifth galaxy (counting from lowest redshift), as they yield the effective redshift $z \simeq 0.29$ and $H(z) \simeq 70$ km/s/Mpc. For this value the uncertainty is in the order of 80\%. To obtain 20\% uncertainty, as \cite{simon2005constraints}, the age uncertainties must be lowered to $\simeq 3\%$. 

The observant reader will notice that the third and fourth galaxies are used for both the first and second Hubble parameter values, meaning that the results are not independent. However, this correlation is not quantified by \cite{simon2005constraints}.

For the $H(z)$ value at the highest redshift we find no combination of galaxies which yields the same effective redshift as the one presented by \cite{simon2005constraints}.

\subsection{The galaxy age uncertainties}
The above analysis demonstrates that a significant reduction in galaxy age uncertainties is necessary to obtain the same precision in $H(z)$ as \cite{simon2005constraints}. It is claimed that some of the systematic errors present in the absolute age determinations cancel when computing the relative ages \citep{Verde}. However, as we argue below, we find that the assumed size of the cancellation underestimates the relative age uncertainties.

First, it is unclear to what degree error bars on the galaxy ages in Fig.~\ref{fig:simonEtAlDataAndResults} (their fig.~1) represent statistical or systematic uncertainties. The ages are determined using the SPEED model \citep{jimenez2004synthetic} from which one obtains a likelihood surface over metallicity and age. Marginalising over metallicity, the age and its \emph{statistical} uncertainty is obtained. Furthermore, systematic errors are discussed after the analysis in their fig.~1 is complete, indicating that systematic effects are ignored, not cancelled.

\begin{figure*}
    \centering
    \includegraphics[width=0.95\textwidth]{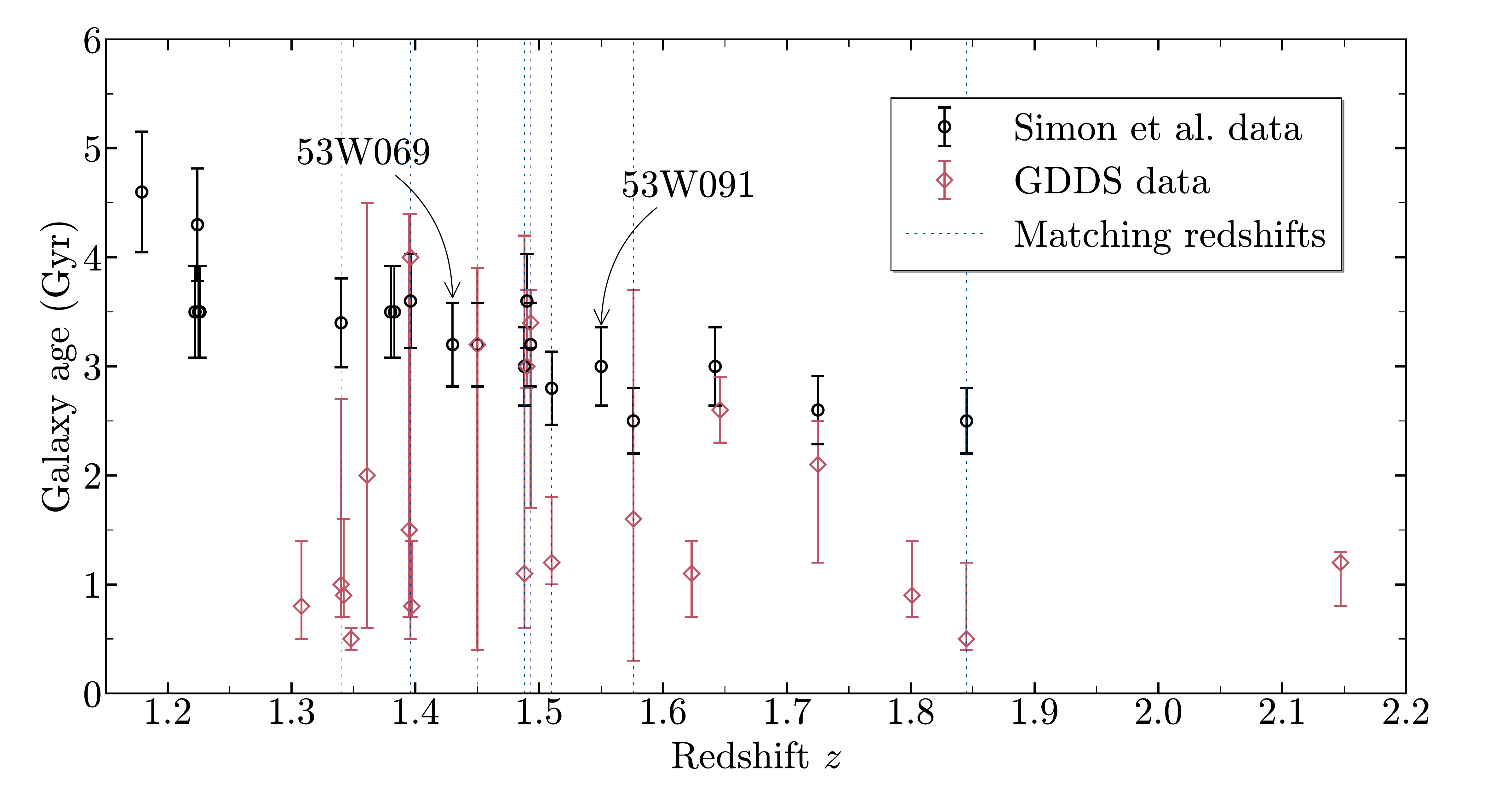}
    \caption{Comparison between the data in \protect\cite{simon2005constraints} and the data presented in \protect\cite{mccarthy2004evolved} for the GDDS data set. Dashed blue lines indicate where the data sets have points with the same redshift. The two galaxies 53W069 and 53W091 are also included.}
    \label{fig:SimonAndGDDScomparison}
\end{figure*}

\begin{figure*}
    \centering
    \includegraphics[width=0.95\textwidth]{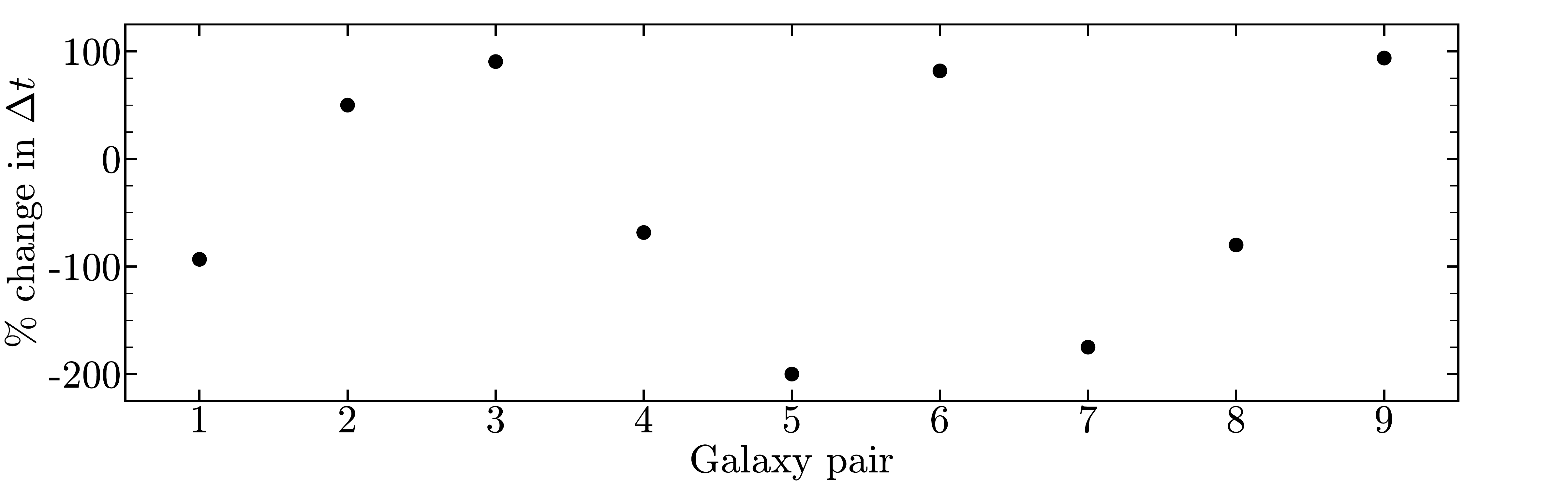}
    \caption{The percentage changes in $\Delta t$ when going from the GDDS data set in \protect\cite{mccarthy2004evolved} to the data set in \protect\cite{simon2005constraints}. Here, $\Delta t$ is only computed for galaxies adjacent in redshift where matching redshifts exist, marked with dashed blue lines in Fig.~\ref{fig:SimonAndGDDScomparison}. At these redshifts, both data sets presumably refer to the same galaxies. Galaxy pairs are counted from lowest redshift, i.e.~pair 1 is computed with the galaxies at the lowest and second lowest redshifts, pair 2 with the galaxies at the second and third lowest redshifts, and so on. }
    \label{fig:deltaTpercChange}
\end{figure*}

Since galaxy age determinations are quite model dependent, we find it unlikely that such a large portion of the uncertainties cancel. 20 of the 32 galaxies in \cite{simon2005constraints} are taken from the Gemini Deep Deep Survey (GDDS) \citep{mccarthy2004evolved} in which ages are estimated. These galaxy ages are reanalysed by \cite{simon2005constraints} using the SPEED model. It is claimed that the ages do not change by much, but Fig.~\ref{fig:SimonAndGDDScomparison} shows that this is not the case, and that using a different model can greatly change the estimated ages. The relative ages may also differ drastically, as seen in Fig.~\ref{fig:deltaTpercChange}, which shows the percentage change in $\Delta t$ when going from the GDDS data set to the data set in \cite{simon2005constraints}. We compute $\Delta t$ for galaxies adjacent in redshift where matching redshifts exist, marked with dashed blue lines in Fig.~\ref{fig:SimonAndGDDScomparison}, as both data sets presumably refer to the same galaxies at these redshifts.

The fact that obtaining the precise $H(z)$ in \cite{simon2005constraints} requires unrealistically small age uncertainties is corroborated when considering that other papers which also compute the Hubble parameter using galaxy ages have one thing in common: many more galaxies are used for each $H(z)$, without obtaining significantly smaller uncertainties. To name a few examples: 1 $H(z)$ is obtained with 7\% precision using $\sim$14000 galaxies by \cite{moresco2011constraining}, 8 $H(z)$ are obtained with 5--17\% precision using $\sim$11000 galaxies by \cite{moresco2012improved}, 2 $H(z)$ are obtained with 21 \& 27\% precision using more than 30 galaxies by \cite{moresco2015raising}, 5 (or 1) $H(z)$ are obtained with 11--16\% (or 6\%) precision using $\sim$130000 galaxies by \cite{moresco20166}, and 1 $H(z)$ is obtained with 75\% precision using 16 galaxies by \cite{ratsimbazafy2017age}.

Furthermore, we simulate an idealised case where galaxy ages are placed on the $\Lambda$CDM prediction for $t(z)$ and galaxy redshifts are placed equidistant between $z=0$ and $z=0.25$.\footnote{Assuming that the data set contains galaxies with $z\in[0,2]$ and one wants 8 $H(z)$ this is the bin width to use.} We find that $\sim70$ ($\sim 280$) galaxies with 12\% age uncertainties are necessary in this bin to obtain $H(z)$ with 20\% (10\%) precision. For bins at higher redshifts even more galaxies are needed.

\section{Reanalysing the data}
We have shown that the resulting Hubble parameter values in \cite{simon2005constraints} are too precise given the employed data. In the following sections we use the same data and reanalyse it with two different methods, Monte Carlo sampling and Gaussian processes, to show specifically how one can obtain $H(z)$ from $t(z)$ and  what uncertainties should be expected using this data set.

\subsection{Monte Carlo sampling}
We place the galaxy data in bins with width $\Delta z$. In each bin we sample the ages from normal distributions and perform a least squares linear fit where the slope is identified as $\dif t / \dif z$. With equation \eqref{eq:HubbleparaFromDerivative} we obtain a distribution for $H(z)$ from which we identify the Hubble parameter as the median and the uncertainty limits by the central region containing 68.27\% of the data. 

In each bin a constant derivative is assumed, which introduces an error. This error is quantified using the $\Lambda$CDM prediction for $H(z)$ as the fractional error $H(z_\mathrm{max})/H(z_\mathrm{min}) - 1$, with $z_\mathrm{max}$ ($z_\mathrm{min}$) being the maximum (minimum) galaxy redshift in the bin. Cosmological parameters for the $\Lambda$CDM model are taken from \cite{PDGreview}.

\begin{figure*}
    \centering
    \includegraphics[width=0.95\textwidth]{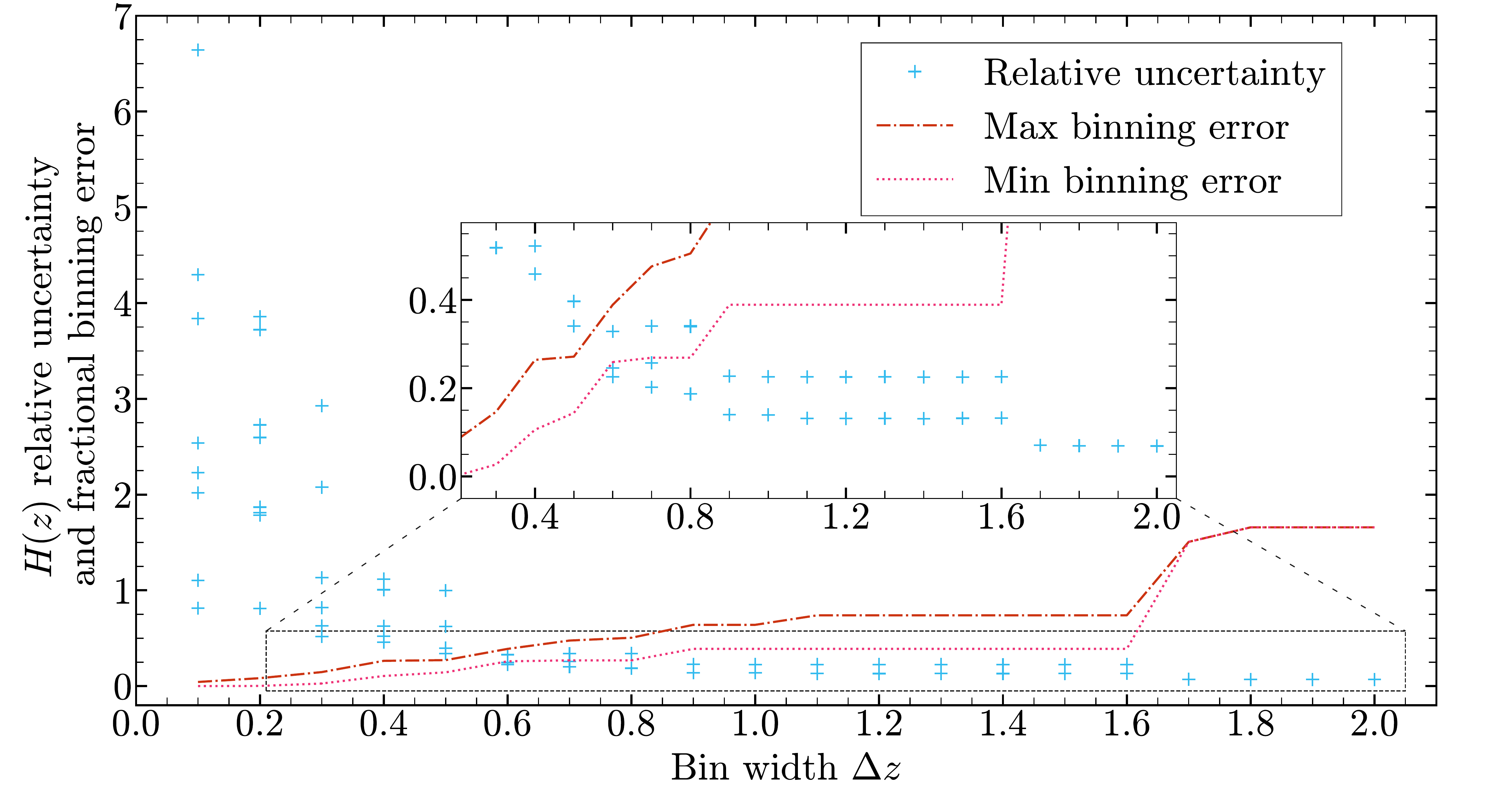}
    \caption{Relative uncertainty in $H(z)$ and the fractional binning errors vs bin width, obtained through Monte Carlo sampling. Relative uncertainty defined as half the width of the 68.27$\%$ area divided by the median of the samples.}
    \label{fig:mcHofZuncVsBinWidth}
\end{figure*}

We present the results in Fig.~\ref{fig:mcHofZuncVsBinWidth}. When 8 $H(z)$ are obtained the uncertainties are larger than $\sim 90\%$. To obtain $\sim 20\%$ uncertainties much larger bin widths must be used, such that only 2 $H(z)$ are obtained. However, then the binning error is too large for the inference to be useful. 

\begin{figure*}
    \centering
    \includegraphics[width=0.95\textwidth]{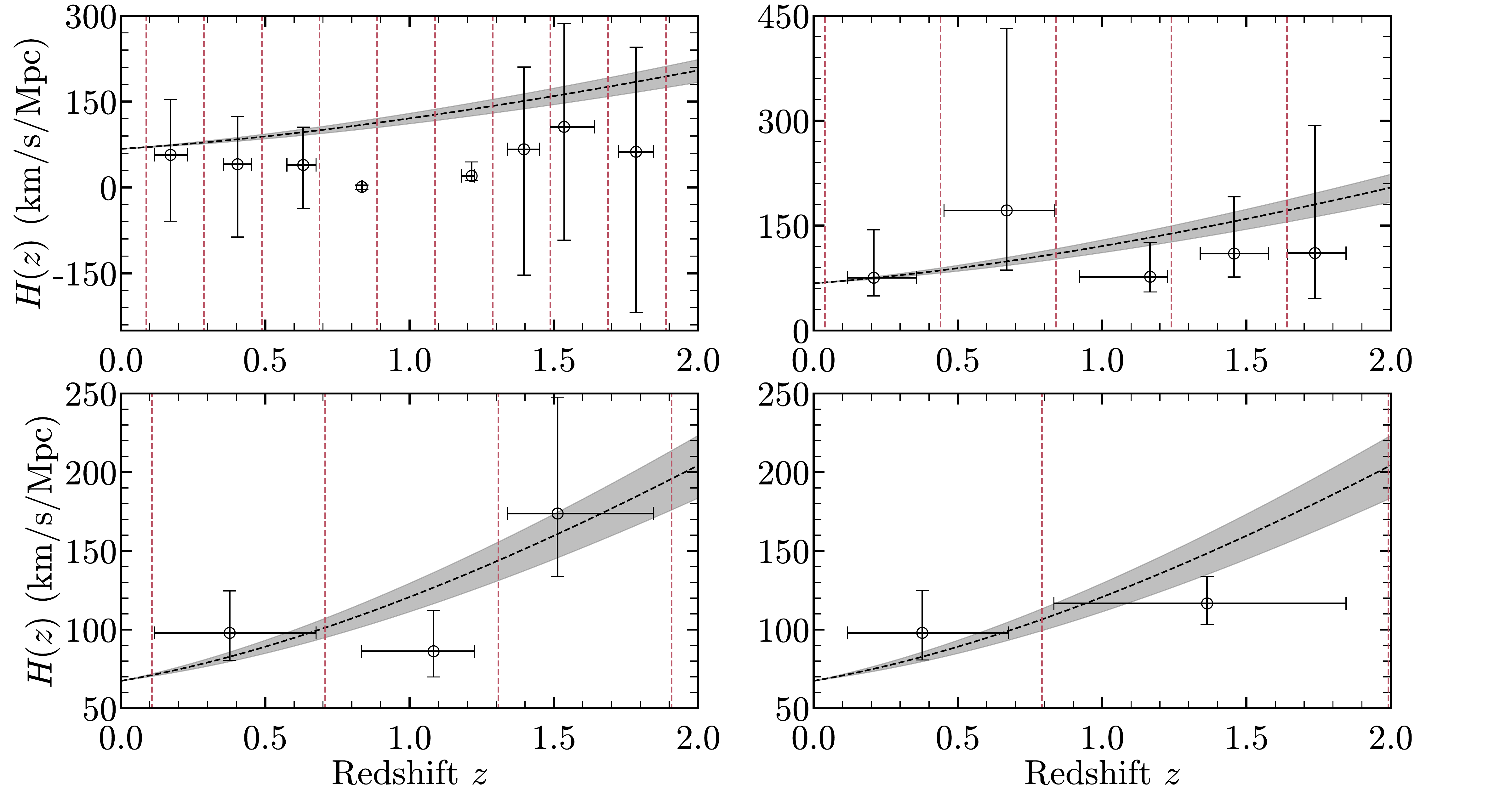}
    \caption{$H(z)$ obtained using Monte Carlo sampling for a few bin widths, with vertical error bars showing the central 68.27\% limits. The dashed black line is the $\Lambda$CDM prediction, and the grey area is obtained by changing $\Omega_\Lambda$ by $\pm 10\%$ and using $\Omega_\mathrm{M} = 1 - \Omega_\Lambda$. In no case is the result good enough to distinguish between different values of $\Omega_\Lambda$ in this range. The dashed vertical lines mark the bins, whereas the horizontal error bars indicate the minimum and maximum redshift of galaxies in the bins. \newline \emph{Top left:} $\Delta z = 0.2$. \emph{Top right:} $\Delta z = 0.4$.
    \emph{Bottom left:} $\Delta z = 0.6$.
    \emph{Bottom right:} $\Delta z = 1.2$.}
    \label{fig:HofZWithLambdaCDMomegaLpm10perc_borders_horizontalerrorbars}
\end{figure*}

To visualise the precision and accuracy of the $H(z)$ obtained we show the actual values in Fig.~\ref{fig:HofZWithLambdaCDMomegaLpm10perc_borders_horizontalerrorbars} for a few select bin widths. For comparison we include the $\Lambda$CDM prediction and vary $\Omega_\Lambda$ by $\pm 10\%$. The resulting $H(z)$ are not precise enough to determine $\Omega_\Lambda$ with this precision and are unlikely to be useful when constraining other cosmological parameters. Furthermore, when 8 $H(z)$ are obtained we can not confidently conclude that the universe is expanding, as the vertical error bars extend to negative values. 

\subsection{Gaussian processes}
The method of Gaussian processes (GPs) generates a continuous inference where every inferred point uses all of the data to some extent. The trade-off is that the inferences at different points are correlated. This method has been used in cosmology for instance to infer properties of dark energy \citep{holsclaw2011nonparametric, seikel2012reconstruction, wang2017improved, zhang2018gaussian, seikel2013optimising}, the Hubble constant\footnote{Note that many papers use the resulting Hubble parameter values from \cite{simon2005constraints} in their inference of $H_0$.} \citep{gomez2018h0, liao2019model, busti2014evidence, bengaly2020hubble}, and other applications \citep{haridasu2018improved, shafieloo2012gaussian, bengaly2020evidence, belgacem2020gaussian}. Thus, we consider this method to infer the Hubble parameter.

A Gaussian process is completely determined by its mean function and covariance function, both of which have to be chosen. We follow convention and choose the zero mean function. As for the covariance function we use five different ones: the squared exponential (SE), and four versions from the so called Matérn class functions. 

For an extensive overview of Gaussian processes we refer to \cite{rasmussen2003gaussian}. However, in this work we follow the method outlined in section 2 of \cite{seikel2012reconstruction}, more specifically section 2.2 and 2.3 where the derivative of a function is reconstructed and then used to construct a function of said derivative by sampling from a predicted distribution of the derivative. We reconstruct $\dif t/\dif z$ and use equation \eqref{eq:HubbleparaFromDerivative} to obtain $H(z)$. 

\begin{figure*}
    \centering
    \includegraphics[width=0.95\textwidth]{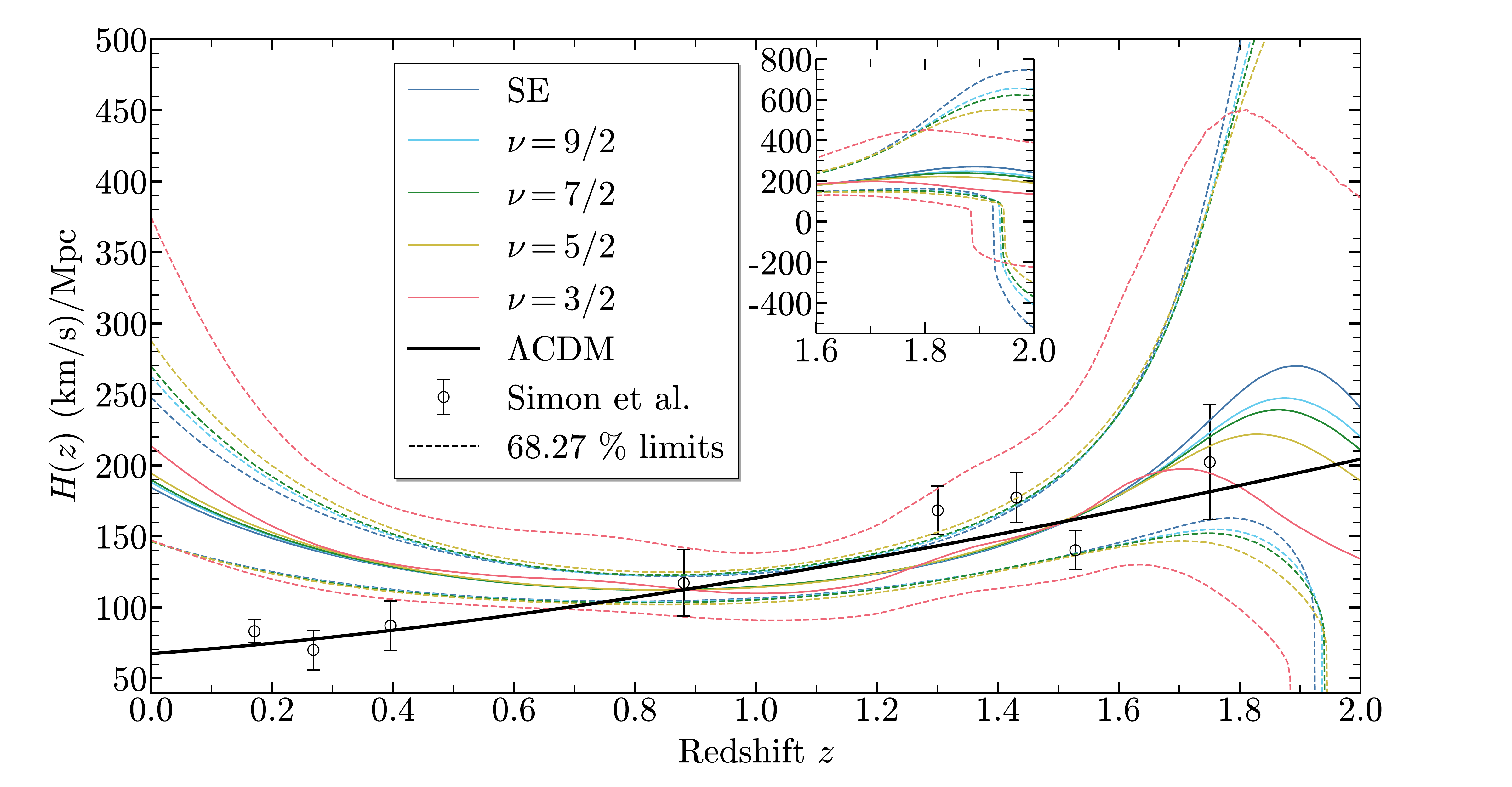}
    \caption{$H(z)$ obtained by sampling from the predicted distribution of $\dif t/\dif z$. Connected lines are the median of the distribution at each redshift $z$. The central 68.27\% of the samplings are within the dashed lines. Also included are the results from \protect\cite{simon2005constraints} and the $\Lambda$CDM prediction for $H(z)$. Inset shows a larger range in $H(z)$ at the higher redshifts.}
    \label{fig:GPHofZmedian68percLimits}
\end{figure*}

\begin{figure*}
    \centering
    \includegraphics[width=0.95\textwidth]{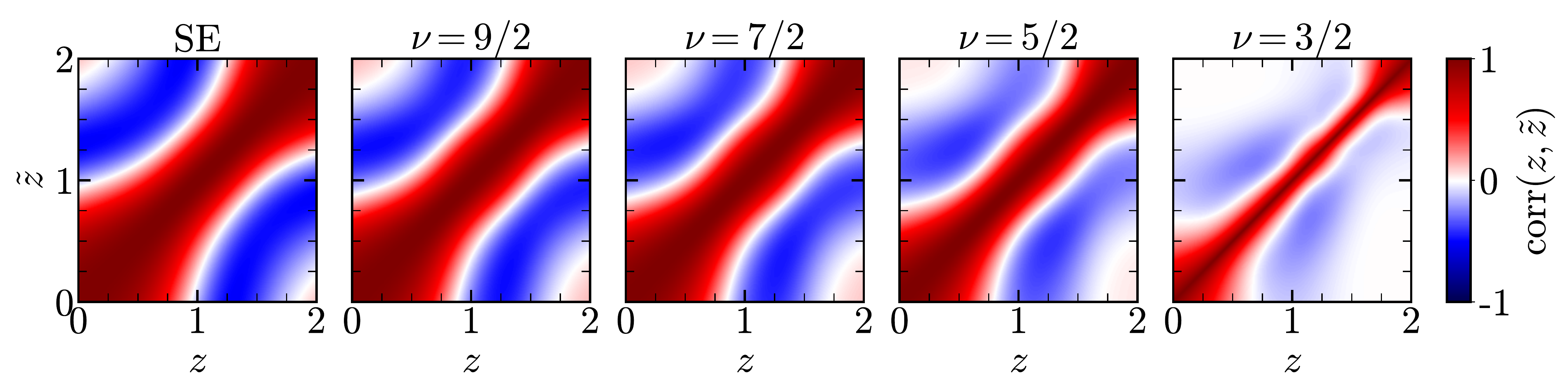}
    \caption{Correlation matrices for the derivative $\dif t/\dif z$, similar to those for $H(z)$. SE refers to the squared exponential, and $\nu$ refers to the corresponding Matérn class function.}
    \label{fig:GPcorrelationsOptimization}
\end{figure*}

The results are shown in Fig.~\ref{fig:GPHofZmedian68percLimits}, where the results from \cite{simon2005constraints} and the $\Lambda$CDM prediction are included for comparison. The uncertainties are very large for larger redshifts and relatively large for the smallest redshifts, but here the GPs divert from the $\Lambda$CDM prediction. For the centre redshifts the GP uncertainties are of comparable magnitude to the ones in \cite{simon2005constraints}. However, considering Fig.~\ref{fig:GPcorrelationsOptimization} we see that these smaller uncertainties come at the cost of strong correlations for inferences at nearby redshifts and quite strong anti-correlations for separated inferences. 


\section{Summary and conclusion}
We have reanalysed the work by \cite{simon2005constraints} in which galaxy age and redshift data is used to obtain the Hubble parameter by computing relative galaxy ages. We have shown that the results are too precise given the precision in the galaxy ages. To obtain the same precision, age uncertainties have to decrease from 12\% to 1--3\%, which is unlikely even if some of the systematic errors in the absolute age determinations cancel when computing the relative ages. Fig.~\ref{fig:deltaTpercChange} indicates that systematic effects between different relative age determinations are in the order of 90\%. Simulating an idealised scenario we found that many more galaxies are needed, where the number is more in line with the number of galaxies used in other papers. Furthermore, some galaxies in \cite{simon2005constraints} are used to compute multiple $H(z)$ meaning that the results are not independent.

Comparing the data in \cite{simon2005constraints} with the cited data we found a quite large difference. Nevertheless, we chose to ignore this discrepancy and reanalysed the data presented in \cite{simon2005constraints}. This was done with two methods: Monte Carlo sampling, and Gaussian processes. The former method showed that obtaining 8 $H(z)$ yields uncertainties larger than $\sim 90\%$. By increasing the bin width 20\% uncertainties could be obtained, but then only 2 $H(z)$ were obtained and the error introduced by the binning was too large for the inference to be useful. The latter method yielded smaller uncertainties for much of the redshift interval, but this came at the cost of strong correlations and anti-correlations. 

The analysis done in this work raises doubts about the accuracy of the resulting Hubble parameter values and corresponding uncertainties by \cite{simon2005constraints} and for this reason, we do not recommend their use for constraining cosmological models.

\section*{Data Availability}
The data in this article is derived from a source in the public domain, where the data being analysed is available in the public domain as figures made by \cite{simon2005constraints} or in a table composed by \cite{samushia2010constraints}.


\bibliographystyle{mnras}
\bibliography{references} 



\appendix
\section{Full reproduction attempt}\label{appendix:Reproducing}
\begin{figure*}
    \centering
    \includegraphics[width=0.95\textwidth]{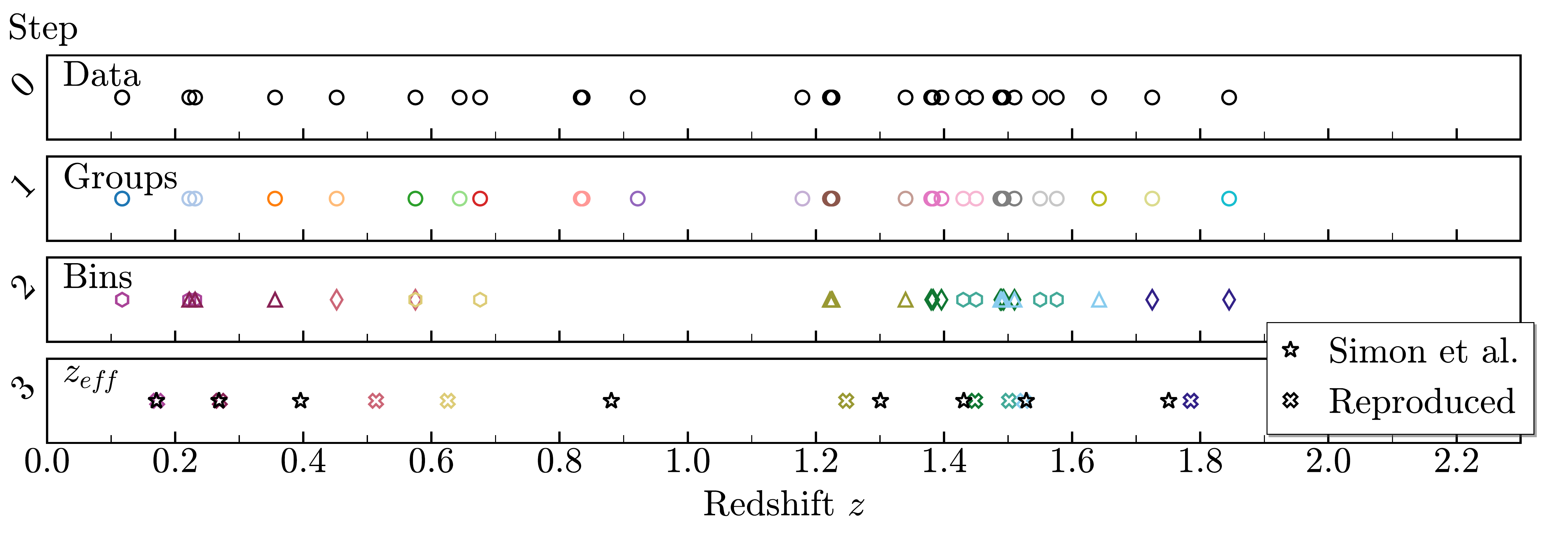}
    \caption{Visualisation of our attempt of reproducing the binning in \protect\cite{simon2005constraints}. \emph{Step 0:} Unprocessed redshift data for the 32 galaxies. \emph{Step 1:} Galaxies within $\Delta z = 0.03$ have been grouped together. Different colours correspond to different groups. \emph{Step 2:} Groups from the previous step which are separated by at least 0.1 but no more than 0.15 have been placed in the same bin. Data points with the same shape and colour correspond to the same bin. Note that some data points are in two different bins. \emph{Step 3:} The effective redshifts in \protect\cite{simon2005constraints} (black and marked with stars), and the effective redshifts computed for every bin from the previous step (coloured as the bins in the previous step and marked with crosses).}
    \label{fig:effectiveRedshiftReproduced}
\end{figure*}

We begin by attempting to reproduce the binning by following the instructions in \cite{simon2005constraints}. We separate the description into three steps, which are visualised in Fig.~\ref{fig:effectiveRedshiftReproduced} (step zero is the unprocessed redshift data for the galaxies in their sample), where the colour coding is described in the figure caption. 

The first step is to simply group together galaxies that are within $\Delta z = 0.03$. After this step it is said that most groups contain more than one galaxy, but we obtain nineteen groups with only eight groups containing two or more galaxies. Nevertheless, in the next step galaxies which are $>2\sigma$ away from oldest galaxy in a group are supposed to be discarded. We find no such galaxies and therefore do not include this step in Fig.~\ref{fig:effectiveRedshiftReproduced}.

The second step is to compute the relative ages for groups which are separated by $\Delta z \in [0.1, 0.15]$. This effectively merges the groups into the final bins, and is what we visualise in Fig.~\ref{fig:effectiveRedshiftReproduced}. 

For each bin obtained we compute the effective redshift as the mean of the redshifts in the bin. This is step three in Fig.~\ref{fig:effectiveRedshiftReproduced}, where we also include the resulting effective redshifts in \cite{simon2005constraints} for comparison. We obtain a quite different set of redshifts and for this reason do not compute the Hubble parameter using the obtained bins. The difference is already large enough to conclude that we find it difficult to reproduce the results by following the instructions. 

\subsection{Matching the effective redshifts}\label{sec:MatchingtheEffectiveRedshift}
As we could not reproduce the binning by following the instructions in \cite{simon2005constraints} we will in this section disregard it and instead attempt to find the bins which yield the same effective redshifts as those presented in \cite{simon2005constraints}. 

\begin{figure*}
    \centering
    \includegraphics[width=0.95\textwidth]{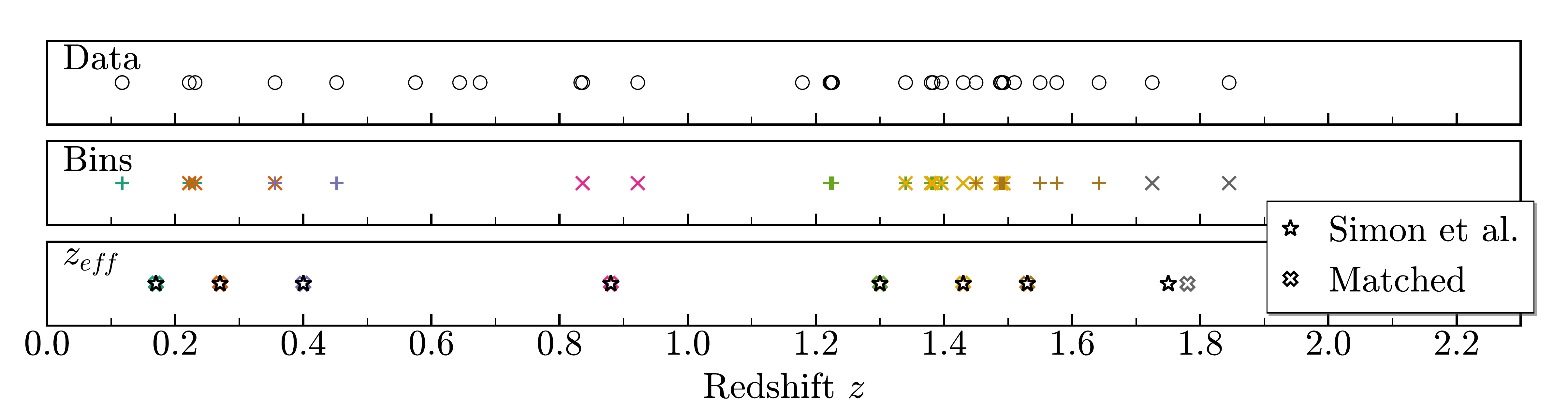}
    \caption{\emph{Top:} Unprocessed redshift data. \emph{Middle:} The bins which produce (almost) the same effective redshifts as in \protect\cite{simon2005constraints}. Data corresponding to the same bin have the same colour. Note that some data points are used twice, i.e~are in two bins. \emph{Bottom:} The effective redshifts of the above bins, with the same colour coding. Only the last redshift was not managed to be matched.}
    \label{fig:matchedEffectiveRedshifts}
\end{figure*}

To this end we first construct all subsets of the 32 data points which have a size ranging from two to ten data points, and for which the difference between the minimum and maximum redshifts is smaller than $0.2$. This limit is chosen since it is slightly more generous than than the limit of $\Delta z = 0.15$ combined with the grouping criteria of $\Delta z = 0.03$, discussed above.\footnote{The reason we do not set it to $0.18$ is because we do not want to be too restrictive, as there is something which we do not understand in binning process in \cite{simon2005constraints}, as demonstrated in the previous section.} After this filtering, the effective redshifts are computed to two decimals for all remaining subsets, defined as the arithmetic mean of the redshifts in each subset. As a last step we compare the obtained effective redshifts with the ones in \cite{simon2005constraints} and choose the bins with matching effective redshift.\footnote{In the case of multiple matches, we choose the bin which produces the closest effective redshift, using all decimals, among the bins with the most amount of data points.} The resulting bins are shown in the middle panel of Fig.~\ref{fig:matchedEffectiveRedshifts}. We are able to match all redshifts, except for the last where we instead choose the bin producing the closest effective redshift.

\begin{figure*}
    \centering
    \includegraphics[width=0.75\textwidth]{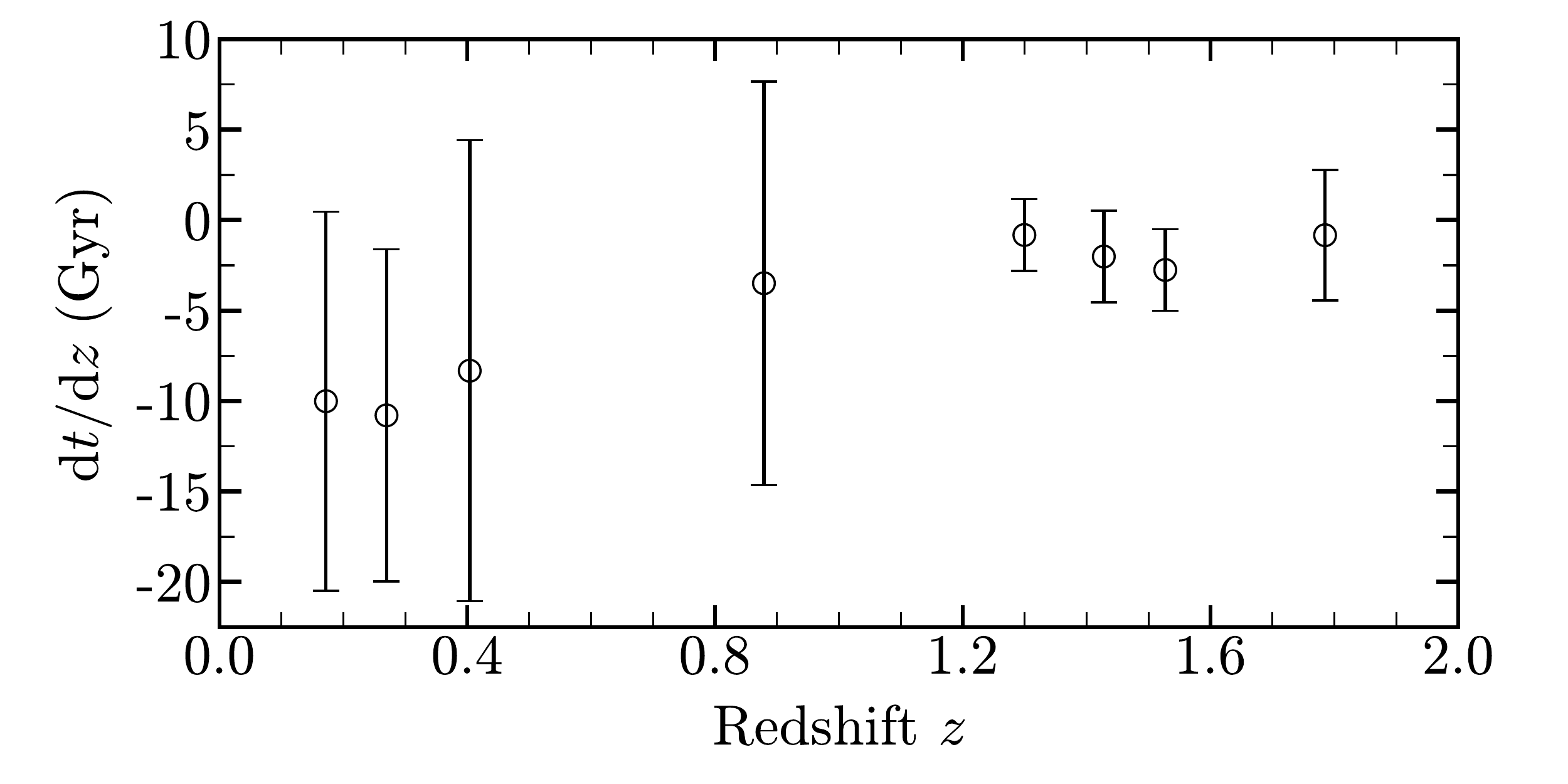}
    \caption{The derivative $\dif t/\dif z$ for the bins which match the effective redshifts in \protect\cite{simon2005constraints}, obtained using a linear $\chi^2$ minimisation fit.}
    \label{fig:dtdzMatchedBins}
\end{figure*}

Using these bins we compute the derivative $\dif t /\dif z$ using a linear $\chi^2$ minimisation fit in each bin. The result is shown in Fig.~\ref{fig:dtdzMatchedBins}, which demonstrates uncertainties in the order of 100\%. The bins used here should be credible to be close to the ones used in \cite{simon2005constraints}, but the obtained uncertainties are significantly larger.


\bsp	
\label{lastpage}
\end{document}